\documentclass[letter,longauth,traditabstract]{aa}%

\newcommand{\hii}{H\,{\scriptsize II}}

\newcommand{\CII}{[C\,{\scriptsize II}]}

\newcommand{\OI}{[O\,{\scriptsize I}]}

\newcommand{\pillars}{{\sl pillars}}
\newcommand{\globules}{{\sl globules}}

\newcommand{\globule}{{\sl globule}}
\newcommand{\Spitzer}{{\sl Spitzer}}

\newcommand{\microm}{{$\mu$m}}
\usepackage{epsfig,graphicx,color}

\usepackage{txfonts}

\begin{document}
\bibliographystyle{aa}
\title{Globules and pillars seen in the [CII] 158 $\mu$m line with SOFIA}  
  \author{N. Schneider\inst{1}
  \and R. G\"usten \inst{2}
  \and P. Tremblin \inst{1}
  \and M. Hennemann \inst{1}
  \and V. Minier \inst{1}
  \and T. Hill \inst{1}
  \and F. Comer\'on \inst{3}
  \and M.A. Requena-Torres \inst{2}
  \and K.E. Kraemer \inst{4}
  \and R. Simon \inst{5}
  \and M. R\"ollig \inst{5}
  \and J. Stutzki \inst{5}
  \and A.A. Djupvik \inst{6}
  \and H. Zinnecker \inst{7,8,9}
  \and A. Marston \inst{10} 
  \and T. Csengeri \inst{2}
  \and D. Cormier \inst{1}
  \and V. Lebouteiller \inst{1}
  \and E. Audit \inst{1}
  \and F. Motte \inst{1}
  \and S. Bontemps \inst{11,12}
  \and G. Sandell \inst{7}
  \and L. Allen \inst{13}
  \and T. Megeath \inst{14}
  \and R.A. Gutermuth \inst{15}
} 
 \institute{
  IRFU/SAp CEA/DSM, Laboratoire AIM CNRS - Universit\'e Paris 
  Diderot, 91191 Gif-sur-Yvette, France
  \and 
  Max-Planck Institut f\"ur Radioastronomie, Bonn, Germany 
  \and
  ESO, Karl Schwarzschild Str. 2, 85748, Garching, Germany 
  \and 
  Boston College, Institute for Scientific Research, MA 02467, USA
  \and 
  KOSMA, I. Physik. Institut, Universit\"at K\"oln, K\"oln, Germany
  \and 
  NOT, 38700 Santa Cruz de la Palma, Spain
  \and 
  SOFIA-USRA, NASA Ames Research Center, Moffett Field, CA 94035, USA 
  \and 
  Astrophysikalisches Institut Potsdam, 14482 Potsdam, Germany
  \and 
  Deutsches SOFIA Institut, Universit\"at Stuttgart, Germany
  \and 
  Herschel Science Centre, ESAC, ESA, Madrid, Spain
  \and
  Univ. Bordeaux, LAB, UMR 5804, F-33270 Floirac, France 
  \and 
  CNRS, LAB, UMR 5804,  F-33270 Floirac, France 
  \and 
  National Optical Astronomy Observatory, Tucson, AZ 85719, USA
  \and
  Dep. of Physics and Astronomy, U. of Toledo, OH 43606, USA
  \and
  Dep. of Astronomy, U. of Massachusetts, Amherst MA USA 
   }



\titlerunning{Globules and Pillars seen with SOFIA}
\authorrunning{N. Schneider}

\date{\today}


\abstract 
{Molecular globules and pillars are spectacular features, found only
in the interface region between a molecular cloud and an
\hii-region. Impacting far-ultraviolet (FUV) radiation creates photon-dominated regions
(PDRs) on their surfaces that can be traced by typical cooling
lines. With the GREAT receiver onboard SOFIA we mapped and spectrally
resolved the \CII\ 158~$\mu$m atomic fine-structure line and the
highly excited $^{12}$CO J=11$\to$10 molecular line from three objects
in Cygnus X (a pillar, a globule, and a strong IRAS source). We focus
here on the globule and compare our data with existing \Spitzer\ data and recent
{\sl Herschel} open-time PACS data. Extended \CII\ emission and more compact
CO-emission was found in the globule. We ascribe this emission
mainly to an {\sl internal} PDR, created by a possibly
embedded star-cluster with at least one early B-star. However,
external PDR emission caused by the excitation by the Cyg OB2 association
cannot be fully excluded.  The velocity-resolved
\CII\ emission traces the emission of PDR surfaces, possible rotation
of the globule, and high-velocity outflowing gas. The globule shows a
velocity shift of $\sim$2 km s$^{-1}$ with respect to the expanding
\hii-region, which can be understood as the residual turbulence of the
molecular cloud from which the globule arose.  This scenario is
compatible with recent numerical simulations that emphazise the effect
of turbulence. It is remarkable that an isolated globule shows these
strong dynamical features traced by the \CII-line, but it demands more
observational studies to verify if there is indeed an embedded cluster
of B-stars.}


   \maketitle


\section{Introduction} \label{intro}

Molecular \pillars\ and \globules\ are visually spectacular features
that are produced when the ionizing radiation from OB-stars forms
elongated structures in the neutral gas at the interface between
\hii-regions and molecular clouds. Though known for a long time, these 
features were popularized by the {\sl Hubble} Space Telescope
(e.g. the famous `Pillars of Creation' in M16). More recent {\sl Herschel}
observations from the HOBYS key program (Motte et
al. \cite{motte2010}) have revealed many nice examples of \pillars\
and \globules\ in Cygnus-X, our SOFIA (Stratospheric Observatory For
Infrared Astronomy) target. An important question is under which
conditions which type of stars can form inside \pillars\ and
\globules. So far, low-mass stars have been found (Sugitani et
al. \cite{sugitani2002}) but to date no clear observational signature
of more massive stars exists. In this {\sl letter} we show
observations that point toward the existence of a cluster of B-stars
associated with a \globule.

\begin{figure}[ht]
\includegraphics[angle=0,width=9cm]{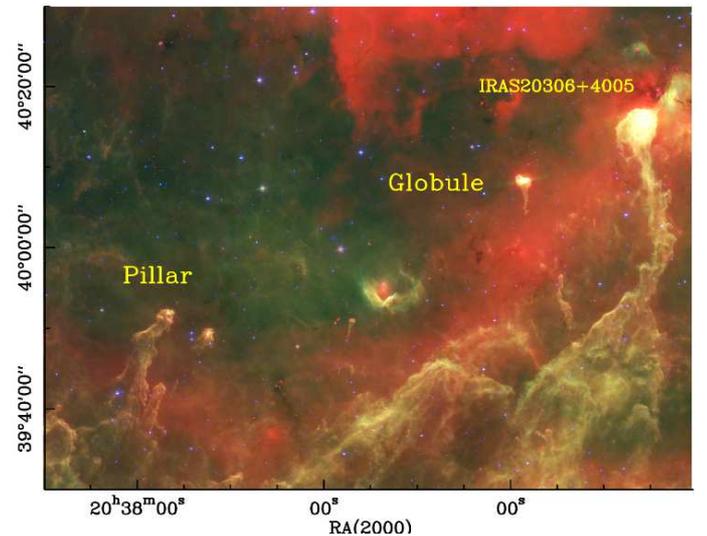}
\caption [] {Three-color (3.6, 8, and 24\,$\mu$m) \Spitzer\ image 
of the southern Cygnus-X region, obtained from the Cygnus-X legacy
survey (Hora et al. \cite{hora2009},
http://www.cfa.harvard.edu/cygnusX).  The Cyg OB2 association is
located at a projected distance of $\sim$1.2$^\circ$ north
($\approx$40 pc, assuming a distance of 1.4--1.6 kpc, Rygl et
al. \cite{rygl2012}). The sources observed with SOFIA (pillar,
globule, IRAS20306+4005) are as indicated.}
\label{overview}
\end{figure}


{\sl Pillars} have a column-like shape and a physical connection to
the gas reservoir of the molecular cloud while {\sl globules} are
isolated and have a head-tail structure pointing toward the
illuminating source (see Fig.~\ref{overview}). Both are only found in
regions of OB-stars where a direct impact of UV-radiation creates
photon-dominated regions (PDRs) on the surface of a molecular
cloud/clump, often visible as a bright rim at the edge of the cloud.
These PDRs are best traced in their cooling lines, i.e. atomic
fine-structure lines of ionized carbon (\CII\, at 158~$\mu$m) and
atomic oxygen (\OI\, at 63 and 145 $\mu$m), as well as high-J CO
rotational lines.

The basic explanation for the formation of {\sl globules}
(e.g. Lefloch \& Lazareff 1994) involves UV-radiation impacting on a
pre-existing clumpy molecular cloud and photoevaporating the lower
density gas, leaving only the densest cores, which may collapse to
form stars (`radiative driven implosion' scenario, Bertoldi
\cite{bertoldi1989}). Recent (magneto)-hydrodynamic simulations including 
radiation (Gritschneder et al. \cite{gritschneder2009}, Tremblin et
al. \cite{tremblin2012a}, \cite{tremblin2012b}) begin to successfully
model {\sl pillar/globule} formation from a different perspective,
emphasizing the importance of turbulence.

To understand {\sl pillar} and {\sl globule} formation, in particular under which
conditions stars can form within them, we initiated a large
observational and numerical study. This program collects observations
from a {\sl Herschel} open time project\footnote{'Pillars of creation:
physical origin and connection to star formation', PI: N. Schneider
(PACS, SPIRE, and HIFI spectroscopy)}, ground-based molecular line
Mopra observations\footnote{Mopra PIs: P. Tremblin, V. Minier (CO,
HCN, HCO$^+$ lines)}, and the SOFIA observations presented here. We
use the HERACLES code (Gonzales et al. \cite{gonzales2007}) and the
radiative transfer PDR-code KOSMA-tau (R\"ollig et
al. \cite{roellig2006}) for modeling.

\begin{figure}
\begin{center}
\includegraphics[width=8.5cm, angle={0}]{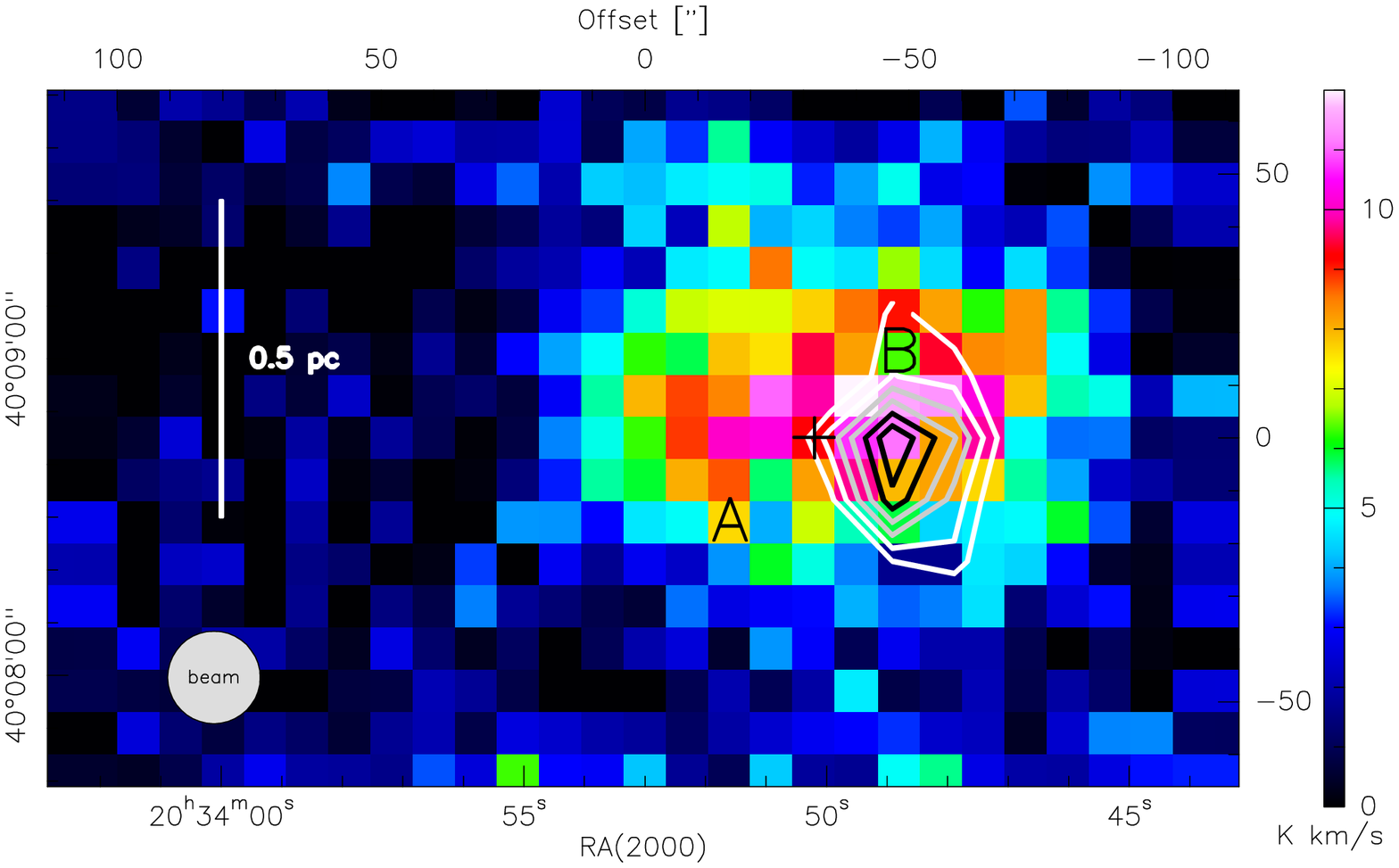}
\includegraphics[width=7.0cm, angle={0}]{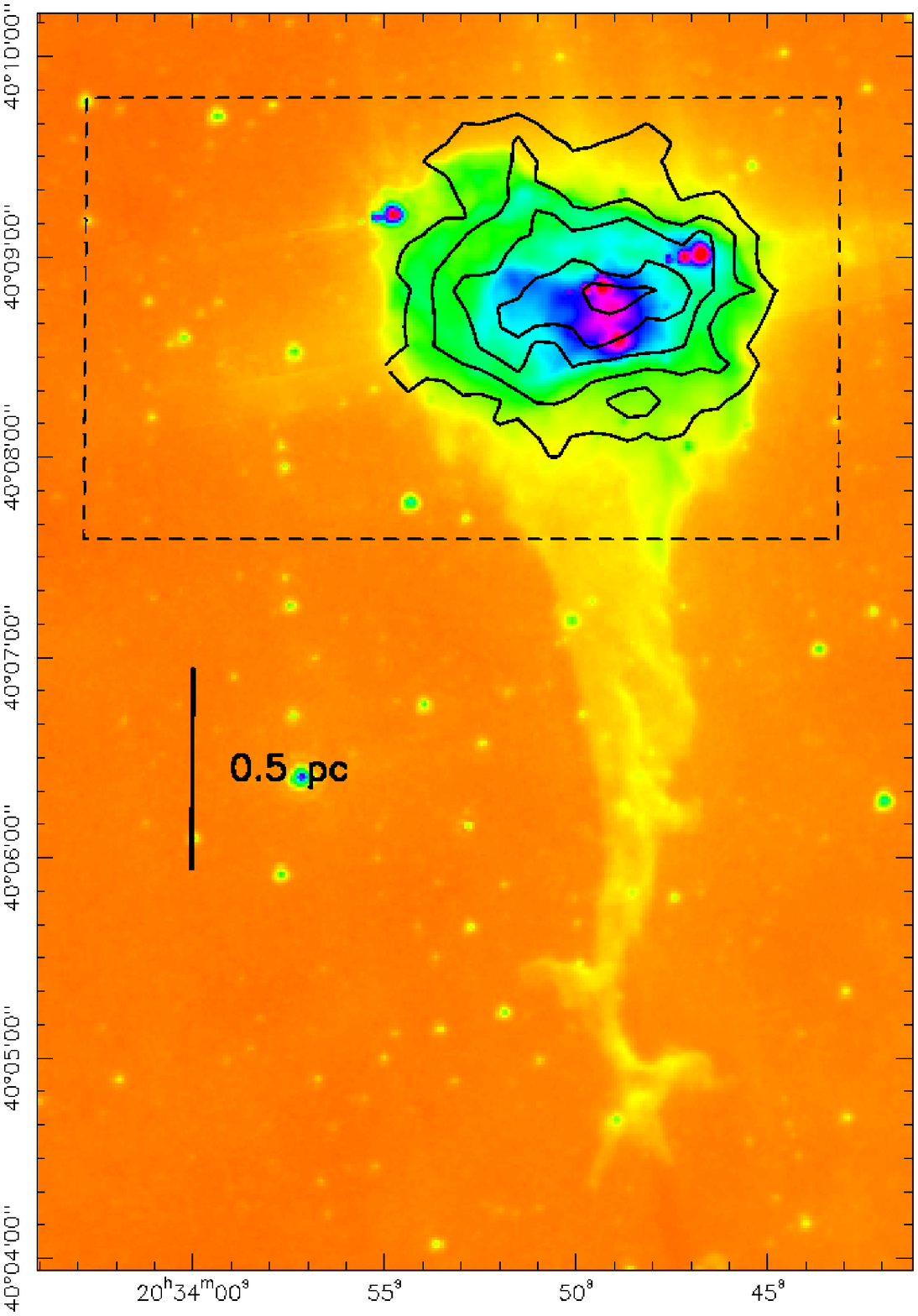}
\end{center} 
\caption []{{\bf Top:} Map of \CII\, emission (color-coded, 
integrated between 3 and 12 km s$^{-1}$ in K kms$^{-1}$ on a T$_{mb}$
scale). The 0,0-position is RA(2000)=20$^h$33$^m$53$^s$
Dec(2000)=40$^\circ$8$'$45$''$. The contours (1 to 3.5 K kms$^{-1}$ by
0.5 K kms$^{-1}$) indicate the very confined CO J=11$\to$10
emission. {\bf Bottom:} Overlay of \CII\, contours (3 to 11 K
kms$^{-1}$ by 2 K kms$^{-1}$ in a T$_{mb}$ scale) in black on
the {\sl Spitzer} 5.8\,$\mu$m continuum emission. The central cluster
and a smaller sub-cluster in the northwest in the head of the globule
(in pink) are clearly visible. The dashed rectangle indicates the area
mapped in \CII. Labels A, B, and the cross indicate the positions of
the \CII\, spectra shown in Fig.~4.}
\label{globule-plot}
\end{figure}

The initial objective of the \CII\, 158~$\mu$m and CO J=11$\to$10
SOFIA observations was to obtain spectrally resolved data to study the
effect of external UV-radiation on \pillars\ and
\globules, and to interpret these in light of numerical simulations. We chose 
to focus on the southern part of the Cygnus-X region (Reipurth \&
Schneider \cite{reipurth2008}) where the very massive and rich Cyg OB2
association (with an estimated 2600 OB-stars, Kn\"odlseder 2000)
illuminates the molecular cloud (see Fig.~\ref{overview}). 


\section{Observations} \label{obs} 

The \CII\, atomic fine-structure line at 1.90 THz (158~$\mu$m) and the
CO J=11$\to$10 molecular rotation line at 1.267\,THz were observed
with the PI-heterodyne receiver GREAT\footnote{The German REceiver for
Astronomy at Terahertz frequencies. GREAT is a development by the MPI f\"ur
Radioastronomie and the KOSMA/Universit\"at zu K\"oln, in cooperation
with the MPI f\"ur Sonnensystemforschung and the DLR Institut f\"ur
Planetenforschung.} on SOFIA on a flight 2011, November 10 from
Palmdale, California.  On-the-fly maps with a scanning speed of
6$''$/s and a size of 4$'\times$3$'$ and 3.5$'\times$1$'$ were made of
the globule and the pillar, and the IRAS source, respectively. The
globule was taken in two coverages, while the pillar and the IRAS
source were observed in only one coverage. The pillar does not have a
sufficient signal-to-noise (S/N) ratio, and the IRAS source is not
fully covered, therefore we did not attempt a more quantitative
analysis of these sources.  We only use a spectrum of each
(Fig. \ref{spectra}) here, because they assist our argument for
globule formation. For blank sky-subtraction, an OFF-positions was
used, based on molecular line and IR-surveys.  A main-beam efficiency
of 0.51 (L2 channel with \CII\,) and 0.54 (L1 channel with CO) was
applied to the data (Heyminck et al. 2012) and a linear baseline was
removed. The observed rms level is 3.3 K (on a main beam temperature
scale in a channel of 0.23 km s$^{-1}$), corresponding to a S/N ratio
of $\approx$\,6 for the peak position. The absolute calibration
uncertainty is estimated to be around 20 \% (Heyminck et al. 2012).

\begin{figure}[ht]    
\begin{center} 
\includegraphics[angle=0,width=7cm]{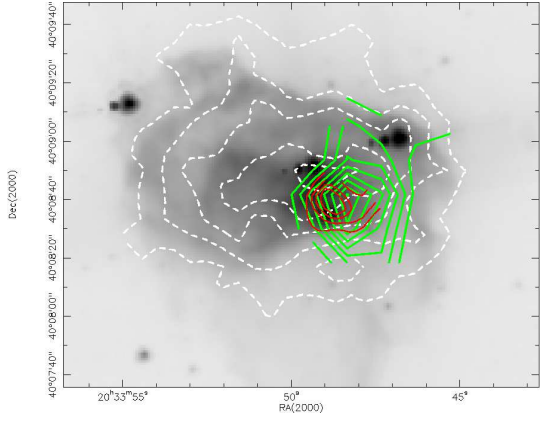}
\end{center} 
\caption [] {
Overlay of PDR-tracer lines on the {\sl Spitzer} 5.8\,$\mu$m continuum
emission. White dashed contours: \CII\ 158 $\mu$m levels 3 to 11 K
kms$^{-1}$ by 2 K kms$^{-1}$. Green contours: CO J=11$\to$10 levels 1
to 3.5 K kms$^{-1}$ by 0.5 K kms$^{-1}$. Red contours: \OI\ 145
$\mu$m 0.7e-16 to 1.5e-16 by 0.2e-16 W/m$^2$.}
\label{herschel}
\end{figure}

In addition, \CII\, 158~$\mu$m, \OI\ 145\,$\mu$m and 63\,$\mu$m data 
from the aforementioned {\sl Herschel} open time project were
obtained for the globule on 2010, December 14, using the PACS
spectrometer (Poglitsch et al. 2010).  These observations consist of a
single pointing (field of view 47$''\times$47$''$) in chopping/nodding
range scan mode, with a chopper throw of 6$'$ and Nyquist spectral
sampling.  The data were reduced using the standard reduction pipeline
in HIPE 7.0. The final maps, from which the line fluxes are measured,
were made with the PACSman software (Lebouteiller et al. 2012).  
These PACS maps (11.5$''$ at 158~$\mu$m) were then convolved
to the approximate beamsize of SOFIA at 158~$\mu$m (16$''$) and the
intensity ratio of individual positions was determined. The SOFIA
\CII\, data are typically a factor 1.6 stronger than the PACS
data. Note, however, that PACS calibration was only derived from 
on-source data (the off-source position was contaminated by emission),
and the calibration for SOFIA is still being tested. Despite this, the
peaks of emission coincide very well, within 5$''$ (the PACS pointing
accuracy is $\approx$2$''$), indicating that the pointing of SOFIA,
using planets and optical pointing with a guide-camera, has been very
reliable. Table~1 summarizes the flux values of each target from both
sets of observations.


\begin{figure}[ht]
\begin{center} 
\includegraphics[width=5.5cm, angle={-90}]{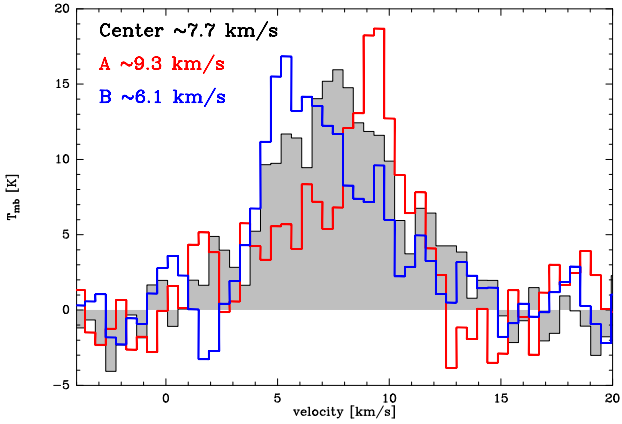}
\caption{ Individual \CII\, spectra from the globule (angular resolution 16$''$), 
with positions A, B, and center (+) as indicated in Fig.~2 (top). The
velocities quoted in the panel are the line-center velocities obtained
by a Gaussian fit. The `center' spectrum has the broadest linewidth
(6.7 km s$^{-1}$) compared to 4.0 and 4.9\,km\,s$^{-1}$ for positions
'A' and 'B'.}
\label{globule-cii}
\end{center}
\end{figure}


\section{Results and analysis} \label{results}

The \CII\ map of the globule and an overlay on \Spitzer\
~5.8\,\microm\ emission is given in Figs. \ref{globule-plot} and
~\ref{herschel}, whilst the
\CII\ spectra at three selected locations are found in Fig. \ref{globule-cii}.
The \CII\ emission covers the head of the globule with an increase in
intensity toward the center. The CO J=11$\to$10 emission is very
confined in the southwest as is also seen in
Fig.~\ref{herschel}, showing a zoom of the globule head and where we
additionally plot contour levels of {\sl Herschel} PACS \OI\ 145 $\mu$m 
emission. All PDR lines tracing in particular {\sl warm} gas are
confined in the same position. We overlay \CII\ and
\Spitzer\ on a near-IR 2MASS image in Fig.~7 where  
both images reveal a number of compact sources that are located at the
peak of \CII\- and CO emission. This is a cluster with at least one
early-type B-star (see appendix A for details). To what extent this
cluster is embedded has a strong implication on the interpretation of
the \CII\ emission distribution.

\begin{figure}[ht]
\begin{center} 
\includegraphics[width=7cm, angle={-90}]{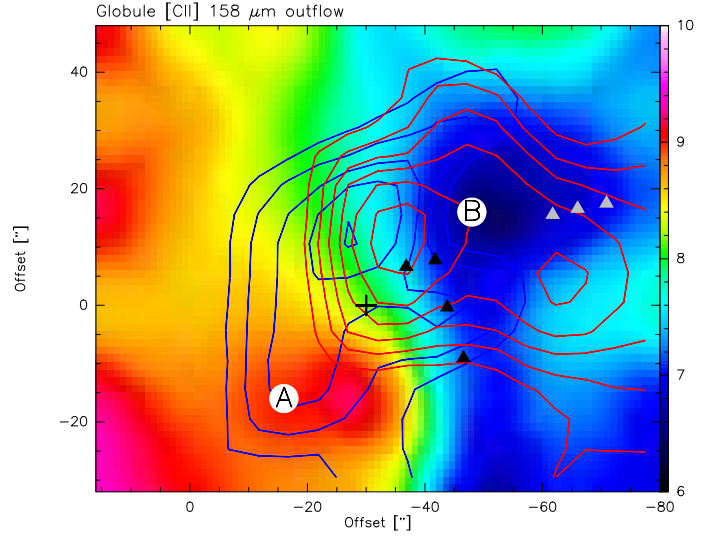}
\caption [] {Close-up of the \CII\, map of the globule with the 
peak line velocity in color, determined from a Gaussian fit to each
spectrum. From each Gaussian fit, the excess emission in the blue
($\sim$4 to 7 km s$^{-1}$) and redshifted ($\sim$9.5 to 11 km
s$^{-1}$) wing was determined and is plotted as blue and red contours
(for better visibility the maps are resampled on a finer grid).  The
black triangles indicate the cluster memebers within the
\hii-region (note that this is where the hot gas, visible in 
CO J=11$\to$10 emission is found). The gray triangles mark the 
three sources from the secondary cluster. 
}  
\end{center} 
\label{outflow}
\end{figure}

Though it is clear, considering the orientation and shape of the
globule, that the impact of the Cyg OB2 association ($\approx$40 pc projected 
distance) created the globule, it is less obvious that the observed
\CII\ emission is only caused by {\sl externally} illuminated PDRs. In
this case, a limb-brightened, arc-like emission distribution with a
decrease of intensity toward the center should be seen. In contrast,
\CII\, \OI\, and CO peak where the brightest sources are found, arguing for
{\sl internal heating} due to the cluster. However, because we do see
2MASS sources with IR excess projected just outside the globule, it is
possible that the globule was larger in the past, formed an extended
cluster, and B-stars from this cluster are now outside the shrinking
globule, illuminating it externally. This may explain why the 
\CII\ emission is quite extended and not as confined as the CO and \OI\ emission. 
On the other hand, beam dilution and exitation by UV-photons leaking
out to the surface with clumps having a small area-filling factor
might also serve as an explanation.

Studying the dynamics of the gas with the \CII\ spectra
(Fig.~\ref{globule-cii}), however, supports the idea that at least
most of the B-cluster is embedded. The `center', located at the
position of the cluster, represents the bulk emission of the globule
around 7.7\,km\,s$^{-1}$ with a broad linewidth, suggesting that
material is moving toward and away from us along the line-of-sight
(LOS). The spectra of positions A and B show a symmetric -- but
opposite -- velocity shift of 1.6--2\,km\,s$^{-1}$ along the LOS. Both
spectra show high-velocity wing emission, i.e. a prominent red wing
for the blueshifted 'B' spectrum and the reverse situation for the 'A'
position.

The velocity field of the central region of the globule is shown in
Fig.~5. There is a sharp border between the two velocity ranges
($\sim$6 and 9 km\,s$^{-1}$), and the star-cluster is spatially
located between these two velocity components.  The velocity field
shows a possible pattern of rotation around a slightly inclined
northsouth axis of the globule, similar to what was already seen in
CO-observations in {\sl pillars} (Gahm et al. \cite{gahm2006},
Hily-Blant et al. \cite{hily2005}). In addition, the globule is
somewhat flattened orthogonal to the rotation axis, which could be
interpreted as an additional argument for rotation. The minimum mass
of the globule needed to support rotation is estimated from $M=({\rm
v}^2 \,r)/ G)$ to 93 M$_\odot$ with a velocity $v$ of 2 km/s and radius $r$ 
0.1 pc, which agrees with mass estimates from dust observations (see
appendix A). Overlaid on the velocity field are blue and red contours of
line-integrated \CII\ emission from the two high-velocity wings,
resulting in a typical `outflow' pattern with two lobes. We attribute
the high-velocity emission to warm gas that is accelerated by the
expanding \hii-region into an inhomogeneous environment. This scenario
is similar to the one found in SOFIA \CII\ observations of S106 (Simon
et al. \cite{simon2012}).

The most probable explanation of the extended \CII\ emission and the
confined CO and \OI\ emission is therefore a combination of internal
heating by the cluster, clumpiness in the globule, and some
contribution from the FUV flux from Cyg OB2. To quantify this
scenario, we determined an FUV intensity of $\chi
\approx$390 at the position of the globule head using the IRAS 60 and 100 $\mu$m
fluxes (Nakagawa et al. \cite{nakagawa1998}). {\sl External
illumination} from Cyg OB2 can contribute to the flux, we estimated a
range between $\chi\approx$180--370, depending on the number of
O-stars (see appendix B for details) but is unlikely. In contrast, {\sl
internal illumination} by at least one B-star produces a flux of
$\chi\approx$360 to 11000 (depending on the spectral type, see
appendix A) in a distance of 0.1 pc from the star, which would be
consistent with our observations.

\begin{figure}[ht]
\begin{center} 
\includegraphics[width=8cm, angle={0}]{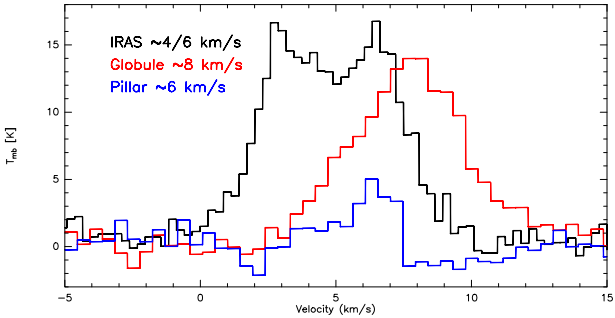}
\end{center} 
\caption [] {Representative \CII\, spectra from the three 
observed regions, the pillar, globule (center position with bulk emission) and 
IRAS20306+4005. The double-peaked spectrum from the IRAS
source is due to two seperate components at around
4 and 6 km s$^{-1}$. The \CII\, spectrum of the globule (red) shows a
shift of around 2 km s$^{-1}$ with respect to the pillar (blue) and
IRAS source (black).}
\label{spectra}
\end{figure} 

\begin{table}[htbp]
 \centering
   \begin{tabular}{lcc}
\hline
\hline
                           & Flux                     &  Flux    \\
                           & [10 $^{-16}$ W m$^{-2}$] &  [10$^{-5}$ erg s$^{-1}$ cm$^{-2}$ sr$^{-1}$]  \\ 
\hline
SOFIA  \CII\, 158 $\mu$m   &  4.34                    &  7.54   \\ 
SOFIA  $^{12}$CO 11$\to$10 &  0.7 (1.6)               &  1.21 (2.78)  \\       
PACS   \CII\, 158 $\mu$m   &  2.71                    &  4.71  \\       
PACS   \OI\, 145 $\mu$m    &  1.24 (1.34)             &  2.15 (2.32)  \\       
PACS   \OI\,  63 $\mu$m    &  2.34 (2.51)             &  4.07 (4.36)  \\       
\end{tabular}
\caption{Flux values from SOFIA and {\sl Herschel} PACS observations of the globule. 
For comparison purposes, all maps have the same angular resolution of
16$''$. The fluxes were determined for the position --40$''$,0 (peak of \CII) though
for some lines, the value of the peak position is offset from this
position (see Fig. 3and is given in parenthesis.}
\label{flux}
\end{table}

\section{Formation of the globule} 

Comparing the \CII\, spectra from the globule, pillar, and IRAS source
(Fig.~\ref{spectra}) shows that the globule has a velocity difference
of $\sim$2 km\,s$^{-1}$ with respect to the other two sources. The
pillar and the IRAS source are still attached to the molecular cloud
(see Fig.~\ref{overview}) but directly face the expanding
\hii\,-region. We therefore define their velocity of $\approx$6
km\,s$^{-1}$ as that of the expanding
\hii\,-region\footnote{This is confirmed by CO 3$\to$2 data that will be 
shown in a forthcoming paper.}. The velocity difference seen for the
globule can then be understood from simulations (Tremblin et al.,
\cite{tremblin2012b}) as the residual turbulence of the molecular
cloud.  When the ionized gas pressure dominates the ram pressure of
the initial turbulence in the molecular cloud, the dynamic is imposed
by the expansion of the \hii\,-region, leading to a global expansion,
with the formation of clumps and pillars in the dense border-shell of
the primordial molecular cloud. When the ram pressure of the
turbulence exceeds the ionized gas pressure, the cold gas has
sufficient kinetic energy to enter the \hii\,-region with motions that
are perpendicular to the ionization propagation.  Then more globules
are formed that show a typical velocity shift with respect to the
cloud of a few km\,s$^{-1}$. The shift can be regarded as the remnant
signature of turbulence inside the molecular cloud.  This picture is
quite different from a radiative-driven implosion scenario of isolated
clumps, which does not predict any particular velocity difference.

\begin{acknowledgements}
Based on observations made with the NASA/DLR Stratospheric Observatory
for Infrared Astronomy. SOFIA Science Mission Operations are conducted
jointly by the Universities Space Research Association, Inc., under
NASA contract NAS2-97001, and the Deutsches SOFIA Institut under DLR
contract 50 OK 0901.

\end{acknowledgements}

\Online

\begin{appendix} 

\section{The IR cluster} 

\onlfig{7}{
\begin{figure}[ht]    
\begin{center} 
\includegraphics[angle=0,width=6.5cm]{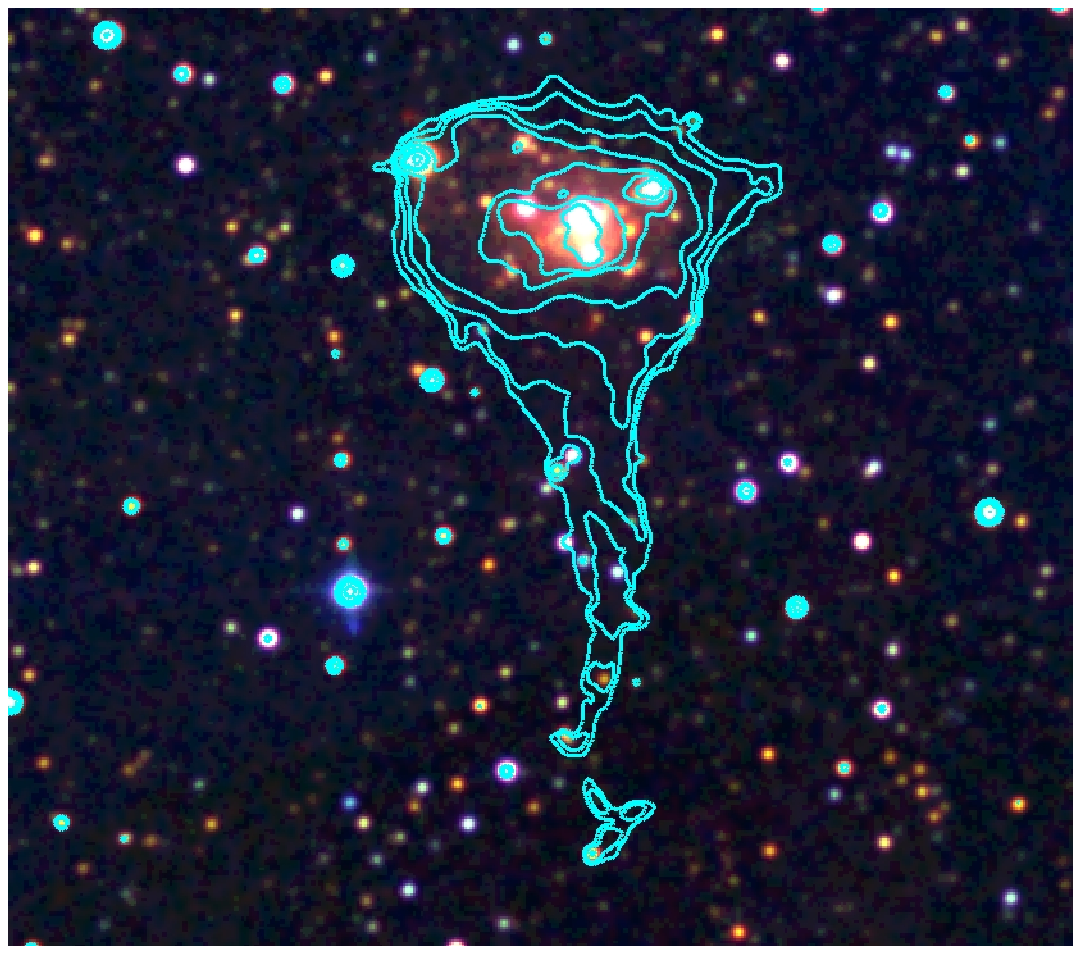}
\includegraphics[angle=0,width=6.5cm]{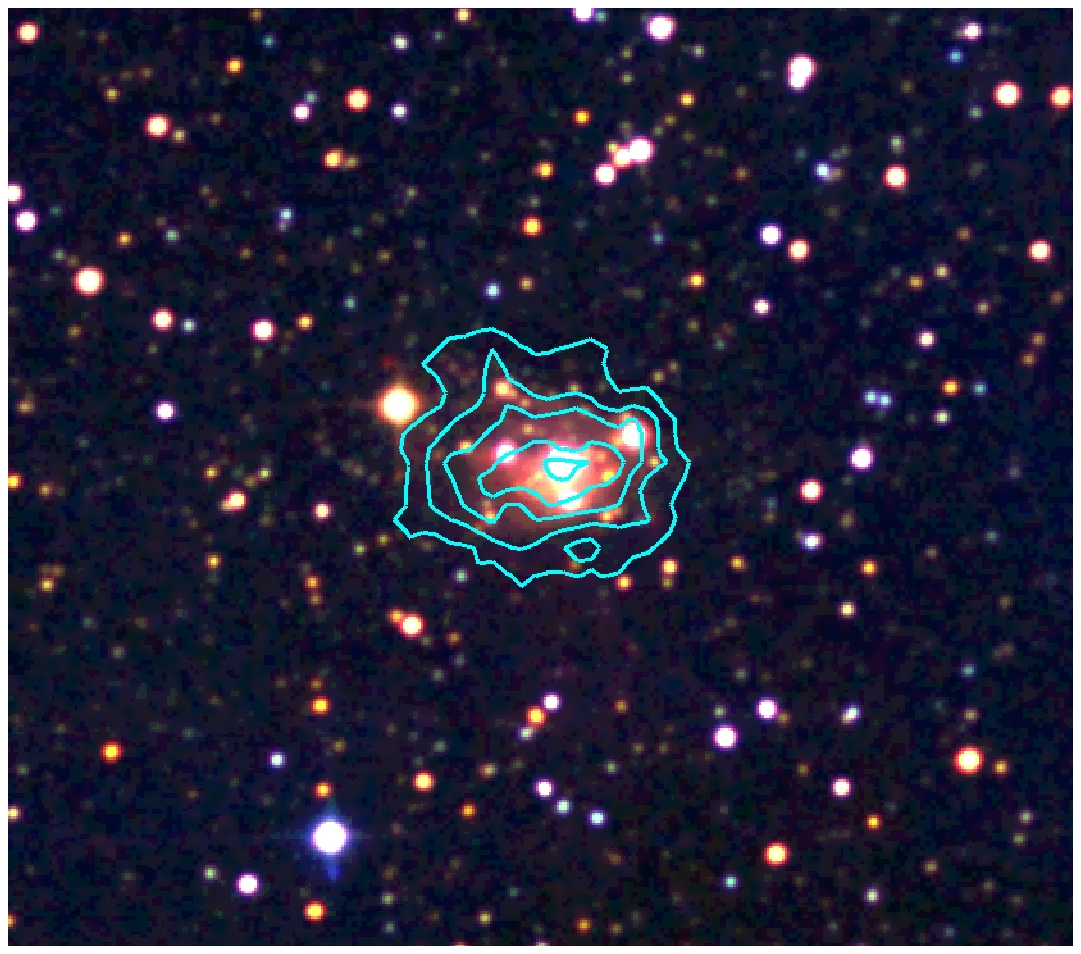}
\end{center} 
\caption [] {
Near-IR image of the 6$' \times$6$'$ field centered on the
cluster in the globule (RGB coded from 2MASS Ks, H and J) with contours 
(28, 30, 35, 50, 125, 200, and 600 MJy/sr) of the \Spitzer\ 5.8 $\mu$m image 
(top) and the \CII\, (bottom) contours (3-11 K km/s in steps of 2 K km/s) overlaid on it. 
North is up, east is left.}  
\end{figure}
}

Figure~7 shows a near-IR image from 2MASS\footnote{The Two Micron All
Sky Survey (2MASS) is a joint project of the University of
Massachusetts and the Infrared Processing and Analysis
Center/California Institute of Technology, funded by the National
Aeronautics and Space Administration and the National Science
Foundation.} with contours of \Spitzer\ and \CII\ overlaid. 
There is a higher stellar density at the location of the
\CII\, emission, coinciding well in size, shape, and position. There is
substantial crowding in the 2MASS images and most of the fainter stars
are typically red. On the red optical plates from the Digital Sky
Survey only a handful of stars are seen. The extended
emission seems brighter in the K-band. Judging from a preliminary J--H vs. H--K diagram,
limited by source-confusion and sensitivity, the central cluster has a
higher fraction of near-IR excess sources compared to the surrounding
field. A preliminary color-color diagram also shows that sources with very
different extinction values in the central cluster are found, though
the median value is about 10 mag of visual extinction of the 2MASS
sources measured. Many of the faint sources detected by 2MASS have no
reliable flux measurement, however, and to make a detailed study of
this cluster we need higher spatial resolution and deeper JHK
images.  

Based on these 2MASS images, both Kronberger et al. (\cite{kron2006})
and Kumar et al. (\cite{kumar2006}) classified the crowding of stars
as 'an associated, partly embedded cluster'. The latter give 9 cluster
members within an effective radius of 0.53 pc, and a stellar mass of
17 M$_\odot$. Optical spectroscopy (Cohen et al. \cite{cohen1989})
identified a B3 III star as one cluster member, but the spectral type
classification is very uncertain because it comes from a rough fit to
Balmer line depths, and the luminosity class III is derived from the
required flux and not from a proper spectral classification.

Sridharan et al. (\cite{srid2002}) derived a
flux of 25\,mJy at 3.6\,cm and 1.4 Jy at 1.1 mm for this star, which
is resolved at both wavelengths, and they interpreted it as a 90
M$_\odot$ clump comprising a high-mass protostellar object and an
\hii-region.  Independently, the spectral type of the ionizing source
can be estimated from the relation between the radio continuum flux
density and the Lyman photon flux (Martin-Hernandez et
al. \cite{martin2005}). A flux of $\sim$13.7 mJy at 1.4 GHz has been
measured (Setia Gunawan et al. \cite{setia2003}), which is consistent
-- as is the flux information at 3.6 cm -- with an ionizing photon
flux of F=1.8$\times$10$^{45}$ ph s$^{-1}$, i.e. a B1 ZAMS star
according to Panagia (\cite{panagia1973}).

%
%
 
Low-angular resolution millimeter continuum observations (BLAST, Roy
et al. \cite{roy2011}; SCUBA, Williams et al. 2005) indicate a dust
temperature of 38 K and a lower limit of the total mass of about 40
M$_\odot$. Williams et al. determined a B2 spectral type from a
luminosity of 4000 L$_\odot$ and dust modeling. Using data from recent
{\sl Herschel} imaging within the HOBYS program (Motte et al.
\cite{motte2010}), we obtained a total mass of 80 M$_\odot$ for the globule (size scale 
$\approx$0.5 pc).  

%
%


Though the spectral classification of the internal B-star is fairly 
uncertain (see above), we can estimate a flux considering different
spectral types. Far-ultraviolet fluxes for a number of early
B-type stars, which include $\alpha$ Vir (B1IV), $\alpha$ Eri (B3V),
and $\beta$ Cen (B1III), were measured by Holberg et
al. (\cite{holberg1982}) using the UV spectrometers onboard the
Voyager interplanetary probes. The spectral types of those three stars
bracket the one estimated for the star at the center of the cluster,
and their FUV luminosities can thus be taken as upper or lower limits
to the UV radiation internally injected in the globule. We used
the distances to these stars as determined by Hipparcos (van Leeuwen
\cite{vanleeuwen2007}) to obtain their luminosities in the 912
\AA\ - 1225 \AA\ range. We obtained luminosities of 1.1$\times$10$^{48}$ 
ph/s ($\alpha$ Vir), 9.7$\times$10$^{46}$ ph/s ($\alpha$ Eri), and
3.0$\times$10$^{48}$ ph/s ($\beta$ Cen). The corresponding
volume-averaged values of $\chi$, adopting a typical radius of 0.2 pc
for the globule, are $\chi$=4,000 ($\alpha$ Vir), $\chi$=360 ($\alpha$
Eri), and $\chi$=11,000 ($\beta$ Cen). These values are higher, or at
least comparable for the lower limit set by the B3V star $\alpha$ Eri,
than the peak value of $\chi$ derived from the IRAS 60 and 100 micron
fluxes, thus supporting an internal origin for the excitation of the
PDR.

\section{Estimation of UV-flux from Cyg OB2}

We here consider only O-stars because they dominate the UV emission over
B-stars, with O6 stars dominating the overall emission output from the
cluster. The number of O-stars in Cyg OB2 is very uncertain and
estimates range between $\sim$45 and $\sim$100 (see Reipurth \&
Schneider \cite{reipurth2008} for an overview). In our estimates, we
assume that $G_0 = 10^6$ at a distance of 0.1 pc from an O-star. The
projected distance from the center of the OB association to the
globule is 40 pc. Hence, 100 O-stars produce a flux of $G_0 \simeq
10^8/400^2=625$ at the position of the globule, corresponding to $\chi
\simeq 370$. We consider this a strict upper limit. For 50 O-stars,
this reduces to $G_0 \simeq 313$ or $\chi \simeq 183$. 

The flux from the IRAS 60 and 100 $\mu$m flux ratio is $G_0 = 660$ or
$\chi \simeq 390$ (obtained with a method described in Nakagawa et
al. \cite{nakagawa1998}) in a large beam at the position of the
globule, which agrees well with the values estimated above. We
believe, however, that the contribution to the FUV flux from the Cyg
OB2 association is significantly lower than the upper limit derived
above due to the reasons detailed below.

Because the volume around the OB association is not empty, there is still
extinction on the way to the globule, which then does not receive the
full UV flux from the entire O-star population. Unfortunately, it is
not possible to give a precise quantification of the effect of
extinction, but a reduction of the estimate of the flux by a factor of
a few is possible.

A basic problem in estimating the number of O stars is the
assumption of a single age for the whole association. Kn\"odlseder
(2000) used 3 Myr and thus assumed that all stars are on the main
sequence, with the brightest stars, therefore, being hottest. Hanson
(\cite{hanson2003}) already showed that some of the stars are bright
because they are (B-type) giants and supergiants, not because they are
hot. There are currently around 70 spectroscopically confirmed O stars
in the association, but more than half of them are O8-O9.

Although it is already difficult to specify a distance between the
O-stars and the globule, the distance of 40 pc to the center of the
association is a {\em projected} distance. Taking into account the
depth along the line of sight can only increase the actual distance,
and because the external flux decreases quadratically with the
distance to the cluster, this in addition may significantly decrease
the FUV flux received by the globule.

\end{appendix}
\end{document}